\documentstyle[epsf]{mn}
\newcommand{\etal}{{\rm et al.~}}
\newcommand{\hmpc}{\;h^{-1}}
\begin{document}

\title[Preferential Clustering Scales]
{A Study of the Large--Scale Distribution of Galaxies in the 
South Galactic Pole Region: II. Further Evidence for a Preferential
Clustering Scale?\thanks{\rm Based on data collected at the European
Southern Observatory, La Silla, Chile.}
}
 
\author[S. Ettori, L. Guzzo, M. Tarenghi]
{S. Ettori$^{1,2}$, L. Guzzo$^{1}$, M. Tarenghi$^{3}$\\
$^{1}$Osservatorio Astronomico di Brera, Via Bianchi 46, I-22055 Merate (CO), 
Italy\\
$^2$Institute of Astronomy, University of Cambridge, Madingley Road, Cambridge CB3 0HA, UK\\
$^{3}$European Southern Observatory, K.~Schwarzschild Str. 2, D-85748 Garching,
Germany}
 
\maketitle

\begin{abstract}

We analyse a set of new pencil--beam galaxy redshift data in three 
small regions around the South
Galactic Pole (SGP) area.  We investigate whether we can find 
any evidence of the quasi--periodic peaks discovered
by Broadhurst \etal (1990) in the distribution of galaxies along the 
NGP--SGP directions.
We use both a power spectrum analysis and a cross--correlation 
with a sliding comb--like window (the comb--template technique).  
Despite the data are less deep ($\sim 600 h^{-1}$Mpc) and certainly
not optimal for such an investigation, there is evidence of the same 
preferential  $\sim 130\; h^{-1}$ Mpc scale in two fields displaced 
respectively $15^\circ$ and $30^\circ$ West of the Broadhurst \etal 
original probe.  Taken alone, however, this scale would not be 
statistically distinguishable from a noise fluctuation.   Nevertheless, 
the statistical significance raises to $\sim 99\%$ when one refers to
the {\it conditional} probability of finding a peak around {\it the same}
scale measured by Broadhurst \etal.

\end{abstract}

 
\section{\bf INTRODUCTION}

Deep ``pencil--beam" surveys (angular size $\sim 1^\circ$, depth $\sim 
10^3 \;h^{-1}$ Mpc) (Koo \& Kron 1987; 
Broadhurst \etal 1988) have produced important 
results about the large--scale structure of the Universe.  They are 
complementary to the wide--angle surveys ($100^\circ $ or larger, 
but with depth only $100-200\; h^{-1}$Mpc; e.g. Giovanelli \& Haynes 1991; 
Strauss \& Willick 1995; Guzzo 1996).  These narrow shots through deep 
space provide 
a confirmation of strong inhomogeneities in the galaxy distribution up
to scales of $50-100 \;h^{-1}$Mpc. 

One of the most exciting results obtained from pencil--beam surveys has been
the detection, along the NGP--SGP axis, of a typical scale in the 
distribution of galaxy concentrations, corresponding to a characteristic
separation of $128 \;h^{-1}$Mpc (Broadhurst \etal 1990, hereafter BEKS). 
This result is consistently confirmed by extensions of the original data 
set (Broadhurst \etal 1995), by independent galaxy surveys in 
nearby areas, as the ESO Slice Project (ESP, Vettolani \etal 1995),
and by the large--scale distribution of clusters of galaxies 
(Bahcall 1991; Guzzo \etal 1992; Tully \etal 1992).  This latter
evidence shows how at least the first maxima do coincide with large--scale
3D structures, and are not artifacts produced by an aliasing effect as
claimed by Kaiser \& Peacock (1990).  However, what the periodicity 
precisely means in terms of large--scale structure is still controversial
(e.g. Dekel \etal 1992).  A similar effect would be expected in
only 15\% of all directions in phenomenological Voronoi models
(e.g. van de Weygaert 1992).

A clearer picture will probably arise only from new pencil--beam
observations in different directions (see Broadhurst \etal 1995), 
together with a detailed knowledge of the 3D power
spectrum of density fluctuations on $\lambda > 100 \hmpc$Mpc.
With this Letter, we try to add a small contribution in this direction,
by analysing some new redshift data in the SGP area.
All through the paper, we use $H_\circ=100\; h$ 
km sec$^{-1}$ Mpc$^{-1}$ and $\Omega_\circ=1$.  

\section {\bf DATA GLOBAL PROPERTIES}

\subsection {\bf The data}

We use the redshift data discussed in Ettori \etal (1995, Paper 1 hereafter),
where full details about the observing strategy and the data reduction 
procedures can be found.   These redshifts were collected using the 33--arcmin
field ESO fibre spectrograph OPTOPUS in three regions around
the SGP, named {\it PW}, {\it PL} and {\it FD}, for each of which we
summarise here the salient aspects.  The completeness of each field 
at different magnitude limits is given in Tab.~1.

\noindent{\bf Field PW} contains 53 redshifts, in the area comprised between
$22^h35^m < \alpha < 22^h40^m$ and $-29^\circ < \delta < 
-27^\circ45'$, limited to $b_j\le 19.5$. 

\noindent{\bf Field PL} contains 415 redshifts, limited to $b_j \le 19.62$.
It extends beetwen $23^h39^m < \alpha < 23^h52^m$ and
$-29^\circ < \delta < -27^\circ45'$, centred on the
rich cluster Klemola 44 (Abell 4038). With a mean velocity
$<cz>\simeq 8800$ km s$^{-1}$, the cluster overdensity dominates 
the redshift distribution in this beam, contributing 
$\sim$ 150 galaxies ($34\%$ 
of the total).  In fact, the main aim of the original survey was 
to study in detail the kinematics and structure of K44 and its surroundings
(Ettori \etal 1996). 
Despite this bias towards a specific redshift peak,
the faint magnitude limit of the survey provides a large number 
of foreground and background redshifts.  To exclude the cluster
effect, we also analyse separately the subsample of 288 galaxies 
laying at $cz > 11500$ km s$^{-1}$, that we call PL--K44. 

\noindent{\bf Field FD} is composed by 5 not--contiguous
Optopus fields (see Paper 1), located between $2^h01^m < \alpha < 2^h16^m$ 
and $-30^\circ10' < \delta < -28^\circ15'$, $\sim 30^\circ$ East of the
BEKS probe, containing 54 galaxies to $b_j\le 19.5$.

\subsection {\bf The Selection Function} 

The comoving distance for all the galaxies in the sample is evaluated as
\begin{equation}
d_c = \frac{c}{H_0 q_0^2 (1+z)}\left\{q_0 z +\left(q_0-1\right)\left[-1+\sqrt{2q_0z+1}\right]\right\},
\end{equation}
with $q_0=0.5$.  The results discussed in the following do not 
change significantly using $q_0=0.1$.  The selection function can be then
calculated by integrating the Schechter luminosity function, 
\begin{equation}
\Phi(L) d(L) =\Phi^*\left(\frac{L}{L^*}\right)^\alpha e^{-L/L^*}d\left(\frac{L}{L^*}
\right),
\end{equation}
to the magnitude limit of each field.  However, in order to take properly
into account the magnitude--dependent incompleteness evident from Tab.~1,
we estimate the actual selection function separately for four 
apparent--magnitude bins.  In other words, we normalize to a different 
$\phi^*$ within each bin, and then calculate the global selection function 
as

\begin{equation}
\phi(d_c) = \sum_1^4 \int_{M_i(b_j,d_c)}^{M_{i+1}(b_j,d_c)} \Phi\left(\Phi^*_i,M\right)
dM.
\end{equation}

For the shape parameters of the luminosity function, we adopt the very recent
estimate from the ESP survey (Zucca \etal 1996), that finds  $M^* = -19.57$ 
and $\alpha = -1.22$ . This estimate of the luminosity function is optimal
for our data, having been performed on same parent photometric catalogue 
(the EDSGC) and to a very similar magnitude limit.  The selection function
also takes into account the K-correction, calculated following Shanks \etal 
(1984), assuming an Early-type/Late-type ratio of 0.3/0.7. This hypothesis 
is necessary, the morphological type being unknown for most of our galaxies.

Fig.~1 shows the resulting $\phi(d_c)$ for the three regions
plotted on top of the observed distribution in bins of 30$\hmpc$ Mpc 
(shaded histogram).   
In the same figures we show also the result of normalizing the data 
to the total number of galaxies expected from the parent photometric 
catalogue, again considering separately the completeness for each 
$b_j$--magnitude bin (open histogram).

\subsection{\bf Data Sparseness}

Before starting to look for possible periodic patterns in our data,
we asked ourselves whether the data were at least good enough to answer
the zero--order question: ``Is the data distribution significantly
different from a random distribution observed with the same selection 
function?''.  One way to quantify this difference is to calculate 
the so--called Double-Root Residuals (DRRs).  This method compares
an observed distribution function with a model by estimating properly 
weighed residuals (Gebhardt \& Beers 1991). Residuals are normally
calculated giving equal weight to both low-populated 
bins, generally belonging to the tails of the distribution, as to central 
bins containing many more data points.   Since the residuals depend strictly 
on $\sqrt{N}/N$ fluctuations, they are not a correct diagnostic of 
the agreement between the data and the model all over the range covered
by the distribution.   The DRR method introduces a square-root trasformation,
which appropriately weighs the residuals of the different bins, showing 
{\it where} there is a significant deviation of the data from the 
model.  If $B_i$ is the data value of bin $i$ and $M_i$ is 
the corresponding model value, the DRRs are defined as: 
\begin{equation}
\left\{ \begin{array}{llll}
        DRR_i & = & \sqrt{2+4B_i} - \sqrt{1+4M_i} & \mbox{if $B_i \geq 1$} \\
        DRR_i & = & 1 - \sqrt{1+4M_i} & \mbox{if $B_i = 0$} 
	\end{array}
\right.
\end{equation}

The local deviations of the DRRs for the given model (assumed to
describe sufficiently well the global trend of the data), give directly
a confidence level in terms of standard deviations.  To reject the null 
hypothesis at the $95\%$ confidence level, for example, we will require 
bins with $|$DRR$| > 2$.

Fig.~2 shows the DRR's for our data, when compared to the selection function. 
For PW and FD the null hypothesis can be rejected at the $95\%$ confidence 
level only for a few bins. This already tells us not to expect to
obtain much significant statistical information from these two data sets.  
On the contrary, PL shows rather significant fluctuations, with peaks in 
corrispondence to K44 and at 170 and 570 $\hmpc$ Mpc and troughs 
around 250, 350 and 450 $\hmpc$ Mpc.

\section{\bf PERIODICITY ANALYSIS}	

Having established that the data show at least some deviations 
from an uniform distribution, we address now the question 
of the presence within the three regions of any significant 
periodic pattern similar to that originally found by BEKS.
To this end, we apply two statistics, the classical Power Spectrum
analysis, and the so--called Comb--Template Test.

\subsection {\bf The Power Spectrum analysis} 

Considering  a set of $N$ data with comoving distance {$d_j$}, we define 
the  {\it power spectrum} as the quantity 
\begin{equation}
P_k = \frac{1}{N}\left| \sum^{N}_{j=1} e^{i2\pi kd_j} \right|^2 .
\end{equation}

This can be seen as the squared amplitude of a discrete Fourier transform of 
$N$ $\delta_{Dirac}$ functions, each located at position $d_j$ 
corresponding to the galaxy positions.  Under the null
hypothesis of a uniform galaxy distribution on scales $> 30 \hmpc$ 
Mpc (Szalay \etal 1991), the probability distribution of $P_k$ is 
equal to a $\chi^2(2)/2$ distribution, i.e. an exponential 
(since the real and imaginary parts of the Fourier
coefficients are Gaussian distributed and statistically independent; cf. 
Burbidge \& O'Dell 1972; see Amendola 1994 for the non-Gaussian case).  
The expectation value of this distribution is unity. To explore 
the significance level of any power $P_k$ present in the spectrum, let us
consider the new variable $z_k = P_k/<P_k>$, where $<P_k>$
is the mean power in the spectrum.  The observed distribution
of $z_k$ will have a mean which is 1 by definition.  We can then evaluate
the significance of any observed power by comparing it to the null hypothesis,
still described by an exponential.

It can be shown (cf. Burbidge \& O'Dell 1972, Szalay \etal 1991) 
that for $K$ independent test wavenumbers
the significance level of any peak $z^h_k$ in the power spectrum  
is given by  

\begin{equation}
p\left(z_k>z^h_k\right) = 1 - \left[1 - \exp (-z^h_k)\right]^K .
\end{equation}
To establish the
significance of a peak, we need therefore to estimate the two parameters 
$<P_k>$ and $K$ for each field. $<P_k>$ is the mean power estimated on the
$K$ wavenumbers; $K$ can be calculated as 

\begin{equation}
K = {k_{max} - k_{min} \over \Delta k},  
\end{equation}
where $k_{max}$ and 
$k_{min}$ correspond respectively to the inverse of the minimum and 
maximum scale that we chose to analyse (see below).  
$\Delta k$ is the spacing in 
wavenumber beetwen two contiguous $k_i$'s, i.e. the smallest independent
frequency and/or the largest comoving range, that for our case 
corresponds to $\Delta k = \frac{1}{d_{j,max} - d_{j,min}} = \frac{1}{700}$.
For our data, we limit the range of investigation to scales between 30 and 400
$h^{-1}$ Mpc, i.e. within a range where we can neglect the effects of 
small--scale clustering (``red noise") 
and of the uncertainty on the selection function
at large distances.   These yield $k_{max}= 0.0333\; h$Mpc$^{-1}$ 
and $k_{min} = 0.0025\; h$Mpc$^{-1}$.

To produce a smoother picture of the power spectrum, we oversampled the
number of wavenumbers, as shown in fig.~3, i.e. calculating the power
at ``dependent'' wavenumbers separated by less than $\Delta k=1/700$. 
It should be clear that this oversampling does not add
any contribution to the significance level of an investigated power.

The final results are shown in fig.~3 and listed in Tab.~2.  
None of the peaks observed in the different fields provides a
statistically significant evidence for a preferred clustering scale.
The probability that the highest peak is simply a random fluctuation
of the noise is everywhere larger than 20\%.   However, it is worth
noticing how for PL--K44 and PW the value of the most probable period is 
interestingly close to the BEKS value of 128 $h^{-1}$ Mpc.

\subsection {\bf The Comb--Template Test}

The Comb-Template Test (CT hereafter), has been used by Duari 
\etal (1992) to look 
for periodicity in the distribution of QSOs. Its main advantage is to
be able to identify, in a simple way, the starting point of a periodicity
(i.e. its phase), 
providing also a relative probability value for the more significant period.

Assuming that the objects are concentrated within peaks of width $w$, spaced
with a period $D$, we slide a comb--like template with periodic ``teeth" 
across the galaxy redshift distribution and evaluate the function 

\begin{equation}
c(d_i,d_\circ) = \left\{ \begin{array}{ll}
                    1 & \mbox{if $\left|d_i - \left\{d_\circ + D\; 
nint\left(\frac{d_i - d_\circ}{D}\right)\right\}\right| <\frac{w}{2}$} \\
                    0 & \mbox{otherwise} 
	\end{array}  \right.
\end{equation}

\noindent where $d_\circ$ is the starting point and $nint(x)$ is the nearest 
integer to $x$. 

For a discrete distribution of $\delta_{Dirac}$ functions, the test consists
in evaluating the correlation function 
\begin{equation}
t(d_\circ) = \sum^{N}_{i=1} c(d_i,d_\circ).
\end{equation}
For an uniform galaxy distribution, it can be shown that $t(d_\circ)$ has the 
a mean value $\mu = Nw/D$ and standard deviation $\sigma_t = 
\sqrt{\mu(1-w/D)}$.
The highest value of $t(d_\circ)$, $t_{max}$, corresponds 
to the most probable starting point $d_\circ$.  In addition, 
given a test period 
the null hypothesis of a uniform distribution can be rejected at the 
$n$-$\sigma$ confidence level 
if $t_{max}$ is larger than $\mu$ by the same amount.
Fig.~4 shows the behaviour of $t(d_\circ)$ for our 4 samples, when the $D$ 
parameter is fixed to the value obtained from the power spectrum analysis 
and the width $w$ is set to 15 Mpc.  We did not 
find any significant dependence of the results on variations of this latter
parameter.  Note that the $d_\circ$ is measured from the lower limit of
the data set in comoving distances, which for the PL--K44 sample 
corresponds to the redshift cut at $cz=11500$ km s$^{-1}$.

A second interesting step is to compare the values of $t_{max}$ obtained 
for different periods, to find which period corresponds to the ``least
random'' distribution.  The results of this test, performed on scale between 
50 and 200 Mpc, are shown in fig.~5.
This analysis is conceptually different from the power spectrum, but,
as can be seen in the figure, it provides a substantial confirmation 
of the results of the harmonic analysis.  Only in one case (PL-K44), there
is some evidence for a different specific scale at $D \sim$ 70  Mpc, which,
interestingly, has a similar counterpart also in  field FD.  We summarise
the results of the CT test  in tab.~3.

\section{\bf DISCUSSION AND CONCLUSIONS}

Let us discuss first the case of the field PL, where the situation is
complicated by the dominating overdensity of the K44 cluster of galaxies.
In fact, the power spectrum (fig.~3, upper panels) shows
a number of peaks, with no particular preference for the BEKS scale.
The main ``period'' which arises from both the power spectrum and the
CT test (fig.~5), clearly corresponds to the separation between
the cluster and the second redshift peak, $\sim 98
\hmpc$Mpc.   The critical effect of the cluster on the harmonic analysis is
also evidenced by the value of the starting point provided by the CT test 
-- i.e. the phase of the hypothetical ``periodicity'' -- which corresponds
exactly to the position of the cluster ($85-86\hmpc$Mpc).
It is interesting to note, however, that when the cluster is removed 
from the sample, a peak arises in $P_k$ right at the BEKS scale, although it
is just slightly more than a $\sim$1--$\sigma$ fluctuation and therefore
not statistically significant. 

The field PW is the only one with a dominant peak in
the power spectrum, corresponding to a period $\sim 120 \hmpc$Mpc.
Also in this case, however, its significance is low, due to the sparseness 
of the data evidenced by the DRRs:   
the null hypothesis can be rejected only at the 77\% confidence level.
The CT test shows the same tendency for a preferential scale around a 
similar value.

In the field FD there is some consistency for a preferred scale at $\sim
70 \hmpc$Mpc from both the $P_k$ and CT analyses.   However, the significance
of this peak is even lower than for the other fields (with a 54\% confidence
level).   We mention it for two reasons: first, because the CT test 
shows a similar scale also for the field PL--K44; second, because 
a similar scale was detected by Mo \etal
(1992) in their analysis of a number of different samples, and is close 
to a kind of second harmonic of the BEKS scale.

So far we have considered the simple absolute significance of the 
potential periods evidenced by our analysis.  In other words, we
have asked ourselves whether in our three sets of data there is
any significant evidence for a regularity with a preferred scale
in the galaxy one--dimensional distribution.  The answer to this
question is clearly negative: we have shown how our set of 
data is essentially not good enough on its own for this purpose.

However, there is a more specific
question that we can ask, i.e. given the three independent data
sets, is there any evidence for a peak in the power spectrum
around the same value found by BEKS, and how significant is it?
The answer to this second question is more interesting, as we
have seen that in two directions, PL and PW, there is indeed
a peak around that same scale.  The significance level of this event
will be now given by the conditional probability of the two
events of (1) having a peak of that height anywhere in the 
explored range, and (2) finding it around $130 \hmpc$Mpc.  In practice,
this results is the product of the absolute probability estimated
in table 2, with the inverse of the number of frequencies explored,
which is 22 in our case.   This gives a probability of 1.4\% for PL
and 1\% for PW (almost 99\% confidence level) that a scale of 
120--130 $\hmpc$Mpc is shown in a sample of galaxies randomly 
drawn from a population of uniform distribution.

It seems difficult to ascribe the systematic detection of this same
scale from many different samples, including the rather sparse one
examined here, to chance coincidences.  While we were writing this
paper, evidence for excess power on 100 $\hmpc$Mpc scale was also 
found in a 2D power spectrum analysis of the Las Campanas Redshift 
Survey (Landy \etal 1996), further supporting the idea that wavenumbers
in the $\sim 100 - 200\hmpc$Mpc range have a particular meaning in the
structuring of our Universe.   One natural possibility is simply that both 1D
(skewers) and 2D (slices) projections amplify and enhance a true maximum
in the three-dimensional power spectrum.  This maximum is expected in
all variants of CDM models, as the signature of the horizon size at the 
epoch of matter--radiation equivalence. Observationally, the spectrum 
has to turnover around these scales, to be able to connect the very 
large--scale estimates from microwave anisotropy measurements 
(e.g. Hu \etal 1996), to the $\sim 100 \hmpc$Mpc--scale direct measurements
from galaxy redshift surveys (e.g. Lin \etal 1996).  To produce the observed
quasi--regular distribution of objects, the spectrum turnover has 
to be rather sharp (Frisch \etal 1995).  Intuitively, the sharper the
3D peak is, the better a well defined preferential cell size is defined.
The next generation of redshift surveys of galaxies and clusters 
(see e.g. Guzzo 1996 for a review), will
provide a first possibility to study both 1D skewers and 3D power spectra
from the same survey volume, thus allowing to finally clarify the
issues discussed here.

\noindent {\bf Acknowledgments.}
\noindent 
We thank G.~Chincarini for his comments suggestions during the early phases
of this work.  SE thanks A. Gaspani for useful discussions on power spectra.

\bigskip\bigskip

\noindent{\bf Figure Captions}
\medskip

{\bf Figure 1.} a) Redshift distribution in 30 $h^{-1}$ Mpc bins in the 
PL field.   The shaded histogram is the observed distribution, while 
the open histogram shows the result of normalizing each apparent magnitude 
bin for the completeness, as shown in Table~1 and discussed in the text.  
The solid curve is the selection function calculated using eq.~3.   
Vertical dashed lines indicate the best periodicity as obtained in chapter 3.
b) The same, for field PW.  c) The same, for field FD.

\medskip
{\bf Figure 2.} The results of the DRRs analysis, showing the significance
of the fluctuations in the data with respect to the null hypothesis
of a uniform distribution with the same selection function. 
The dashed lines at $\pm$ 2 indicate the 95\% significance level.

\medskip
{\bf Figure 3.} Power Spectra for the four data sets analysed.
The vertical dashed lines correspond to the 128 $h^{-1}$ Mpc BEKS periodicity 
($k \simeq 0.0078\;h\;Mpc^{-1}$).

\medskip
{\bf Figure 4.} The distribution of the phases $d_\circ$ from 
the Comb--Template Test.   The period adopted (as obtained from the power 
spectrum), is reported for each data set.

\medskip
{\bf Figure 5.} The distribution of the test periods from the 
Comb--Template Test.  $n$ gives the number of $\sigma$ deviations 
from a uniform distribution, for any given $D$.  The arrows mark 
the highest peak, corresponding to the best period.  The BEKS 
periodicity is given by the dashed lines.

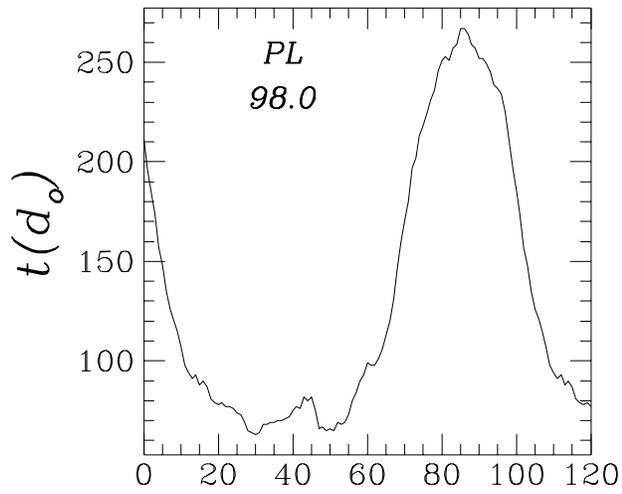
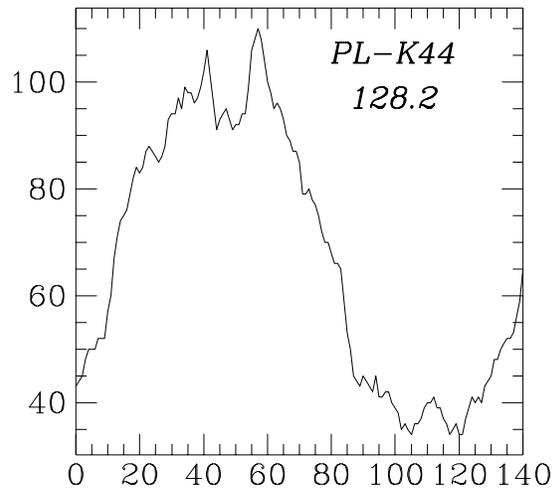
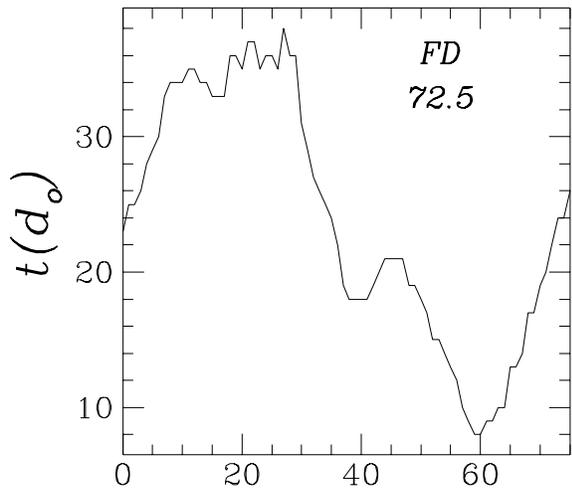
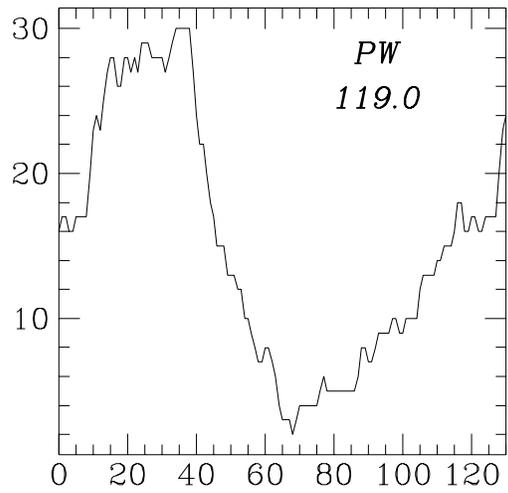
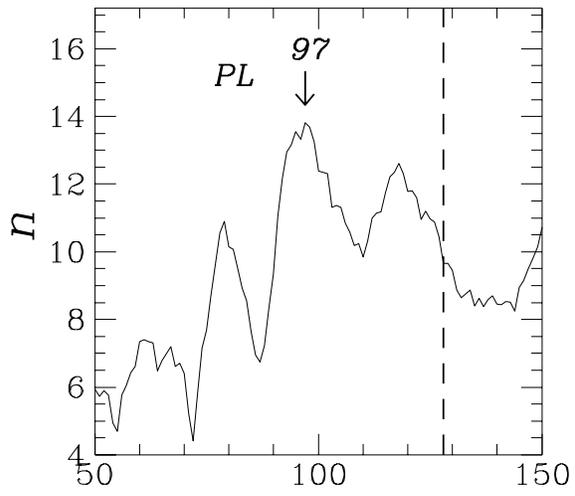
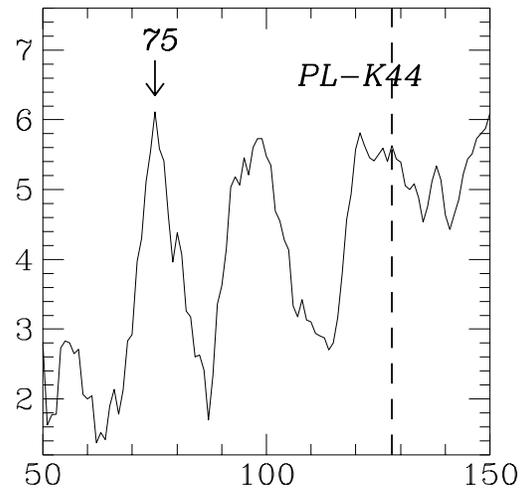
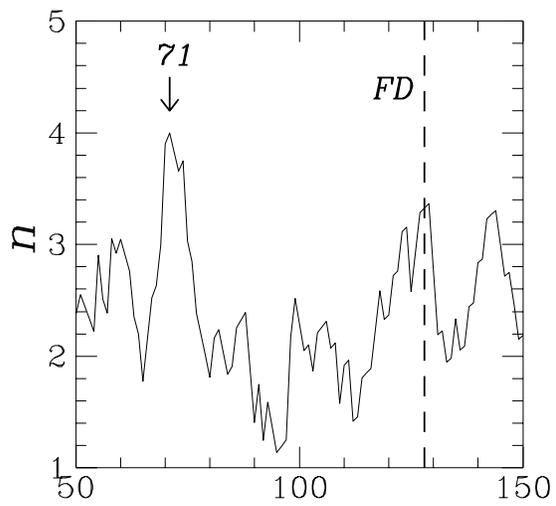
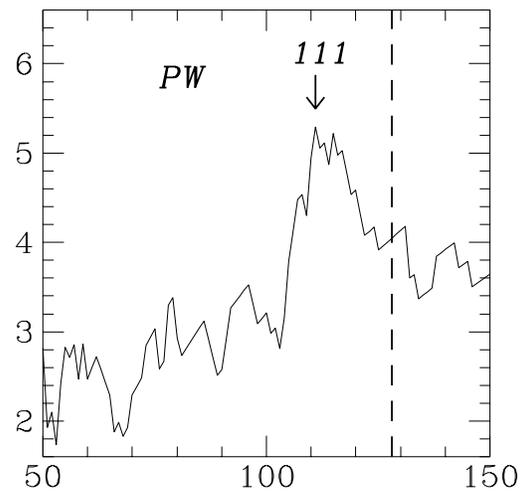
\end{document}